# Thermal analysis for peristaltic flow of nanofluid under the influence of porous medium and double diffusion in a non-uniform channel using Sumudu Transformation Method

**Asha S. K[*] and Namrata Kallolikar[1]**

[*,1]Department of Mathematics, Karnatak University,

E-mail: as.kotnur2008@gmail.com[*]

nckallolikar@gmail.com[1]

*Abstract.* The present study aims to investigate the study of double-diffusive convection onperistaltic flow under the assumption of long wavelength and low Reynoldsnumber. The mathematical modelling for a two-dimensional flow, along withdouble diffusion in nanofluids is considered. The motivation of the presentresearch work is to analyse the effects of thermal radiationon a peristaltic flow through a porous medium in a non-uniform channel.The heat flux of the linear approximation employs the thermal radiation ofthe flow problem. The governing equations are analytically solved by using Homotopy Perturbation Sumudu Transformation method (HPSTM) with the help of the symbolic software Mathematica.The results of the velocity, pressure rise, temperature, solutal(species)concentration and nanoparticle volume fraction profiles are graphicallyshown.

## 1. Introduction

Peristalsis pumping is a unique mechanism and well known to physiologists as a natural mechanism of pumping materials. Peristaltic transport is a form of fluid material transportinduced by a progressive wave of area either by contraction or expansion along the length of a distensible tube. This natural phenomenon is known as peristaltic flow. Peristaltic pump is first coined by Latham [1] in 1966 and further research work on Peristaltic flow was extended by Shapiro et al. [2] and Jaffrin et al. [3] many theoretical and experimental studies have been discussed for the various type of fluid flow channel in Peristalsis motion. Peristaltic flow occurs widely in various physiological functions such as blood flow in small vessels of the human circulator system, semen transport in vas deferens, movement of ovum in the fallopian tube, bile transport from gall bladder to duodenum, ingesting food via the esophagus, chyme movement in the gastrointestinal tract, vasomotion of blood vessels such as veins, capillaries and vertebral arteries, transport of urine from the kidney to the bladder, transport of hygienic fluids, transfer of corrosive fluids, transport of toxic fluids in the nuclear power industry.

Phenomenon of Peristaltic transport in non-uniform ducts may be of considerable interest, it is noted that many physiological problems are known to be of non-uniform cross-section. Gupta et.al [4] and Srivastava et.al [5] have considered peristaltic transport in non-uniform channels.It is seen in many physiological structures that the ducts are either in uniform or non-uniform cross section.It is well known that the human body is made up of several non-uniform ducts,for example lymphatic vessels, intestine, ducts efferent's of the reproductive tract.Some other works are done on theoretical studies of peristaltic transport of physiological fluids in vessels of non-uniform cross section [6-8].

Double diffusion is the phenomenon in which mass and heat transfers occur concurrently with complicity of the fluid motion. Research on peristaltic flow of double-diffusive convection is an innovative concept.Double diffusion has important applications in chemical engineering, geophysics, oceanography, solid-state physics, astrophysicsand also many engineering applications like natural gas storage tanks, solar ponds, metal solidification processes and crystal manufacturing.Peristaltic pumping with double diffusive convection in nanofluid was studied by [9-12].They show double diffusive effect with other nanofluid models.

Effects of thermal radiation and heat generation/absorption on hydromagnetic peristaltic flow is of considerable significance for many scientific and engineering applications viz. heating and freezing of chambers, fossil fuel combustion energy processes, evaporation from massive open water reservoirs, propulsion devicesfor aircraft, missiles, satellites and space vehicles etc. The research work on peristaltic motion with the influence of thermal radiation is reported in References [13-15].They show the effect of some other physical parameter like Hartmann number, thermophoresis and Brownian parameters.

In all abovementioned investigations reveal that the role of thermal radiation effects on double diffusive convection of nanofluid in non-uniform cross-section has not yet been studied in literature. The motivation of the present research work is to analyse the effects of thermal radiation on a peristaltic flow through a porous medium in a non-uniform channel. Double diffusive convection problems have so many practical applications in oceanography, geophysics, biology, and astrophysics. Heat transfer rate is controlling by applying the radiation. The basic governing equations are highly nonlinear, which are solved by using Homotopy perturbation Sumudu transformation method (HPSTM) [16-20] and analysis of the embedded parameters on velocity, pressure rise, energy, solutal (species) concentration, and nanoparticle volume fraction is shown in the form of graphs.

## 2. Mathematical analysis

Here we have considered the two-dimensional peristaltic flow of non-uniform channel with the sinusoidal waves of small amplitudes propagates the speedof the channel walls. $c$ is the constant speed of the channel (Figure 1).

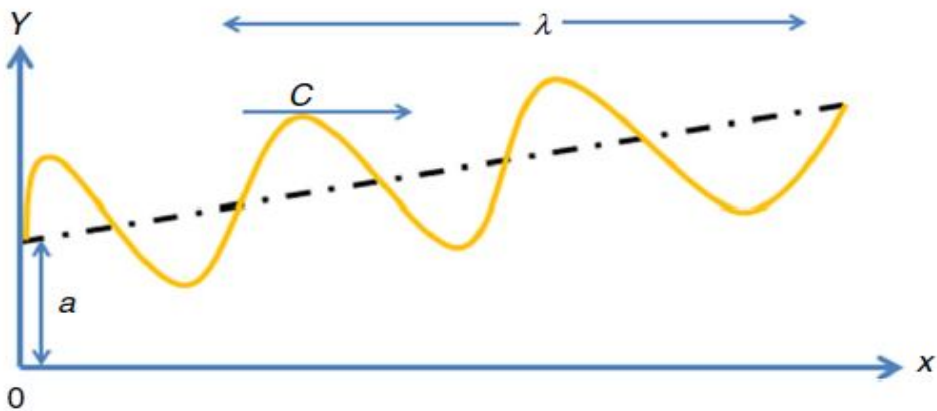

Fig. 1 Geometry of a peristaltic transport through a non-uniform channel

The geometric model of the channel is defined as

$$h'(X,t) = a + d \sin\left[\frac{2\pi}{\lambda}(\bar{X} - c\bar{t})\right] \tag{1}$$

Where $a = a_0 + k\bar{X}$ is the channel half width, $c$ is the constant wave speed, $\lambda$ is the wavelength, $\bar{t}$ is the time and $d$ represents the amplitude of the wave. The velocity components $\bar{U}$ and $\bar{V}$ along the $\bar{X}$ and $\bar{Y}$ directions respectively, in the fixed frame, the velocity field $V$ is taken as

$$V = [\bar{U}(\bar{X},\bar{Y},\bar{t}), \bar{V}(\bar{X},\bar{Y},\bar{t}), 0] \tag{2}$$

The radiative heat flux $q_r$ can be written as

$$q_r = -16 \frac{\sigma^* T_1'^3}{3k^*} \frac{\partial T'}{\partial y'}. \tag{3}$$

In the above equation, $k^*$ is the Rosseland mean absorption coefficient and $\sigma^*$ denotes the Stefan-Boltzmann constant. Considering the nanofluid flow temperature is very small, therefore the term is a linear function of temperature.

The basic governing equations describing the peristaltic flow patterns for the nanofluid are as follows:

$$\frac{\partial U'}{\partial x'} + \frac{\partial V'}{\partial y'} = 0, \tag{4}$$

$$\rho f \left(\frac{\partial U'}{\partial t'} + U' \frac{\partial U'}{\partial x'} + V' \frac{\partial U'}{\partial y'}\right) = -\frac{\partial p'}{\partial x'} + \mu \left(\frac{\partial^2 U'}{\partial x_2'} + \frac{\partial^2 U'}{\partial y_2'}\right) + \rho f g (\varphi' - \varphi_0') \\ + \rho f g (T' - T_0') - g(\rho_p - \rho f_0)(F' - F_0'), \tag{5}$$

$$\rho f \left(\frac{\partial V'}{\partial t'} + U' \frac{\partial V'}{\partial x'} + V' \frac{\partial V'}{\partial y'}\right) = -\frac{\partial p'}{\partial y'} + \mu \left(\frac{\partial^2 V'}{\partial x_2'} + \frac{\partial^2 V'}{\partial y_2'}\right), \tag{6}$$

$$(\rho c)_f \left(\frac{\partial T'}{\partial t'} + U' \frac{\partial T'}{\partial x'} + V' \frac{\partial T'}{\partial y'}\right) = k_T \left(\frac{\partial^2 T'}{\partial x_2'} + \frac{\partial^2 T'}{\partial y_2'}\right) + \frac{D_{TC}\alpha C_p}{C_S}\left(\frac{\partial^2 \varphi'}{\partial x_2'} + \frac{\partial^2 \varphi'}{\partial y_2'}\right) - \frac{\partial^2 q_r}{\partial y_2'} \\ + (\rho c)_p D_B \left(\frac{\partial F'}{\partial x'}\frac{\partial T'}{\partial x'} + \frac{\partial F'}{\partial y'}\frac{\partial T'}{\partial y'}\right) + (\rho c)_p \frac{D_T}{T_m}\left[\left(\frac{\partial T'}{\partial x'}\right)^2 + \left(\left(\frac{\partial T'}{\partial y'}\right)^2\right)\right], \tag{7}$$

$$\frac{\partial \varphi'}{\partial t'} + U' \frac{\partial \varphi'}{\partial x'} + V' \frac{\partial \varphi'}{\partial y'} = D_{CT}\left(\frac{\partial^2 \varphi'}{\partial x_2'} + \frac{\partial^2 \varphi'}{\partial y_2'}\right) + D_s\left(\frac{\partial^2 T'}{\partial x_2'} + \frac{\partial^2 T'}{\partial y_2'}\right), \tag{8}$$

$$\frac{\partial F'}{\partial t'} + U' \frac{\partial F'}{\partial x'} + V' \frac{\partial F'}{\partial y'} = D_B\left(\frac{\partial^2 F'}{\partial x_2'} + \frac{\partial^2 F'}{\partial y_2'}\right) + \frac{D_T}{T_m}\left(\frac{\partial^2 T'}{\partial x_2'} + \frac{\partial^2 T'}{\partial y_2'}\right), \tag{9}$$

where $(\rho c)_f$ is the heat capacity of the fluid, $D_{TC}$ is the Dufour diffusivity, $k_T$ is the thermal conductivity of the fluid, $(\rho c)_p$ is the effective heat capacity of the nanoparticle material, $g$ is the gravity, $\varphi'$ is the solutal

(species) concentration, $K_0$ is permeability constant of the porous medium, and $\sigma$ is electrically conductivity of the fluid. Furthermore, $\rho_f$ is the effective density, $D_T$ is thermophoresis diffusion coefficient, $D_s$ is the solutal diffusivity, $C_p$ is the specific heat at constant pressure, $\alpha$ is the thermal diffusion ratio, $C_s$ is the concentration susceptibility, $D_{CT}$ is the Soret diffusivity, and $T_m$ is the mean fluid temperature.

The connection between the wave frame and laboratory frame are introduced through
$$u' = U' - c, \quad y' = y', \quad v' = V', \quad x' = x' - ct'.$$
(10)

Introducing the following nondimensional variables

$$\left.\begin{array}{l} \psi = \dfrac{\psi'}{ca_o}, \; x = \dfrac{x'}{\lambda}, \; y = \dfrac{y'}{a_o}, \; t = \dfrac{ct'}{\lambda}, \; v = \dfrac{v'}{c}, \; \delta = \dfrac{a_o}{\lambda}, \; p = \dfrac{p' a_o^2}{\mu c \lambda} \\[6pt] u = \dfrac{u'}{c}, \; \alpha = \dfrac{k_T}{(\rho c)_f}, \; \Pr = \dfrac{\mu}{\alpha}, \; b = \dfrac{d}{a_o}, \; N_{TC} = \dfrac{D_{TC} \alpha C_P (\phi_1' - \phi_0')}{\mu k_T C_s (T_1' - T_0')} \\[6pt] \theta = \dfrac{T' - T_0'}{T_1' - T_0'}, \; \varphi = \dfrac{\varphi' - \varphi_0'}{\varphi_1' - \varphi_0'}, \; \gamma = \dfrac{F' - F_0'}{F_1' - F_0'}, \; Nt = \dfrac{(\rho c)_p D_T (T_1' - T_0')}{(\rho c)_f \mu T_m} \\[6pt] Rd = 16 \dfrac{\sigma^* T_1'^3}{3 k^* k_T}, \; h = \dfrac{h'}{a_o}, \; f^* = \dfrac{q}{ca_o}, \; Gr_t = \dfrac{\rho_f g a_o^2 (T_1' - T_0')}{c\mu} \\[6pt] Gr_c = \dfrac{\rho_f g a_o^2 (\varphi_1' - \varphi_0')}{c\mu}, \; N_{TC} = \dfrac{D_{CT}(T_1' - T_0')}{D_s (\varphi_1' - \varphi_0')}, \; \mathrm{Re} = \dfrac{\rho_f c a_o}{\mu} \\[6pt] Gr_F = \dfrac{(\rho_p - \rho_{f_0}) g a_o^2 (F_1' - F_0')}{c\mu}, \; Nb = \dfrac{(\rho c)_p D_B (F_1' - F_0')}{(\rho c)_f \mu}. \end{array}\right\}$$
(11)

where $\delta$ is the dimensionless wave number, $\theta$ is the dimensionless temperature, $\varphi$ is the nanoparticle volume fraction, and the stream function taken as $v = -\delta \dfrac{\partial \psi}{\partial X}$ and $u = \dfrac{\partial \psi}{\partial y}$.

By using Equation (11), Equations (4) to (10) can be written as

$$\frac{\partial p}{\partial x} = \frac{\partial^2 u}{\partial y^2} + Gr_T \theta + Gr_C \varphi - Gr_F \gamma,$$
(12)

$$\frac{\partial p}{\partial y} = 0,$$
(13)

$$\frac{\partial^3 \psi}{\partial y^3} + Gr_T \frac{\partial \theta}{\partial y} + Gr_C \frac{\partial \varphi}{\partial y} - Gr_F \frac{\partial \gamma}{\partial y} = 0,$$
(14)

$$\frac{\partial^2 \theta}{\partial y^2} + Nb \Pr \frac{\partial \theta}{\partial y}\frac{\partial \gamma}{\partial y} + N_{TC} \Pr \frac{\partial^2 \varphi}{\partial y^2} + Nt \Pr \left(\frac{\partial \theta}{\partial y}\right)^2 + Rd \frac{\partial^2 \theta}{\partial y^2} = 0,$$
(15)

$$\frac{\partial^2 \varphi}{\partial y^2} + N_{CT} \frac{\partial^2 \theta}{\partial y^2} = 0,$$
(16)

$$\frac{\partial^2 \gamma}{\partial y^2} + \frac{Nt}{Nb}\frac{\partial^2 \theta}{\partial y^2} = 0.$$
(17)

The corresponding dimensionless boundary conditions can be written in the following form

$$\left.\begin{array}{l}\psi = 0, \quad \frac{\partial^2 \psi}{\partial y^2} = 0, \quad \theta = 0, \quad \gamma = 0 \quad \text{at} \quad y = 0, \\ \psi = f^*, \quad \frac{\partial \psi}{\partial y} = -1, \quad \theta = 1, \quad \gamma = 0 \quad \text{at} \quad y = h = 1 + \frac{\lambda k_o x}{a_o} + b\sin[2\pi(x-t)].\end{array}\right\}$$
(18)

where $f^*$ is the wave frame it is related to the dimensionless time mean flow rate $\Theta$ in the laboratory frame as follows

$$\Theta = f^* + 1, \quad f^* = \int_0^h \frac{\partial \psi}{\partial y} dy,$$
(19)

where $\Theta = \frac{Q}{sa_1}$ and $f^* = \frac{q}{sa_1}$ are the dimensionless time mean flow rate in a fixed and wave frames respectively.

The coupled partial differential Equations (14) to (19) with the boundary conditions (18) are solved by using Homotopy perturbation sumudutransformtion method (HPSTM) and the analysis of the present method of solution is mentioned in the below section.

## 3. Method of solution

The governing equations (12) to (17) are evaluated by HPSTM. This method is an analytical technique, which can be used to calculate the nonlinear problems that contains large and small physical parameters(solution is obtained in terms of convergent series solutions).

In this section, we apply the HPSTM tothe governing equations to obtain an approximate analytical solution. By applying the sumudu transform, inverse sumudu transform on both sides ofthe governing equations we get,

$$u(y) = ay + y\frac{\partial p}{\partial x} - s^{-1}\left[v^2 s\left[Gr_T \theta - Gr_C \varphi + Gr_F \gamma\right]\right],$$
(20)

$$\psi(y) = ay^2 - s^{-1}\left[u^3 s\left[Gr_T \frac{\partial \theta}{\partial y} + Gr_C \frac{\partial \varphi}{\partial y} + Gr_F \frac{\partial \gamma}{\partial y}\right]\right],$$
(21)

$$\theta(y) = ay - s^{-1}\left[u^2 s\left[\frac{Nb\Pr}{1+Rd}\frac{\partial \theta}{\partial y}\frac{\partial \gamma}{\partial y} + \frac{N_{TC}\Pr}{1+Rd}\frac{\partial^2 \varphi}{\partial y^2} + \frac{Nt\Pr}{1+Rd}\left(\frac{\partial \theta}{\partial y}\right)^2\right]\right],$$
(22)

$$\varphi(y) = ay - s^{-1}\left[u^2 s\left[N_{CT}\frac{\partial^2 \theta}{\partial y^2}\right]\right],$$

(23)

$$\gamma(y) = ay - s^{-1}\left[u^2 s\left[\frac{Nt}{Nb}\frac{\partial^2 \theta}{\partial y^2}\right]\right].$$

(24)

Now applying HPM,

$$\sum_{m=0}^{\infty} p^m u_m = ay + y\frac{\partial p}{\partial x} - s^{-1}\left[v^2 s\left[Gr_T \sum_{m=0}^{\infty} p^m H_{1_m} - Gr_C \sum_{m=0}^{\infty} p^m M_{1_m} + Gr_F \sum_{m=0}^{\infty} p^m C_{1_m}\right]\right],$$

(25)

$$\sum_{m=0}^{\infty} p^m \theta_m(y) = ay - s^{-1}\left[u^2 s\left[\frac{Nb\Pr}{1+Rd}\sum_{m=0}^{\infty} p^m H_{2_m}(y) + \frac{N_{TC}\Pr}{1+Rd}\sum_{m=0}^{\infty} p^m M_{2_m}(y) + \frac{Nt\Pr}{1+Rd}\sum_{m=0}^{\infty} p^m \left((\theta_m)_y\right)^2\right]\right],$$

(26)

$$\sum_{m=0}^{\infty} p^m \gamma(y) = ay - s^{-1}\left[u^2 s\left[\frac{Nt}{Nb}\sum_{m=0}^{\infty} p^m H_{3_m}\right]\right],$$

(27)

$$\sum_{m=0}^{\infty} p^m \varphi(y) = ay - s^{-1}\left[u^2 s\left[N_{CT}\sum_{m=0}^{\infty} p^m H_{4_m}\right]\right],$$

(28)

$$\sum_{m=0}^{\infty} p^m \psi_m(y) = ay^2 - s^{-1}\left[u^3 s\left[Gr_T \sum_{m=0}^{\infty} p^m H_{5_m}(y) + Gr_C \sum_{m=0}^{\infty} p^m M_{5_m}(y) + Gr_F \sum_{m=0}^{\infty} p^m C_{5_m}\right]\right],$$

(29)

where, $H_{1_m}, M_{1_m}, C_{1_m}, H_{2_m}, M_{2_m}, H_{3_m}, H_{4_m}, H_{5_m}, M_{5_m}, C_{5_m}$ are He's polynomials. So He's polynomials are given by

$$H_m(U_0, U_1, U_2, \ldots U_m) = \frac{1}{m!}\frac{\partial^m}{\partial p^m}\left[N\left(\sum_{i=0}^{\infty} p^i U_i\right)\right]_{p=0}.$$

(30)

Comparing the coefficient of like powers of p in equation 25 to 29, we get the required approximations and we get the solution as follows

$$u = ay + y\frac{\partial p}{\partial x} - Gr_T a\frac{y^3}{6} - Gr_C a\frac{y^3}{6} + Gr_F a\frac{y^3}{6} + Gr_F \frac{y^3}{6} + \frac{Nb\Pr}{1+Rd}Gr_T a^2 \frac{y^4}{96} + \frac{Nt\Pr}{1+Rd}Gr_T a^2 \frac{y^4}{96} + \ldots$$

(31)

$$\theta = ay - \frac{Nb\Pr}{1+Rd}a^2\frac{y^2}{2} - \frac{Nt\Pr}{1+Rd}a^2\frac{y^2}{2} + \frac{Nb^2\Pr^2}{(1+Rd)^2}a^3\frac{y^3}{6} + \frac{NbNt\Pr^2}{(1+Rd)^2}a^3\frac{y^3}{6} - \frac{Nt\Pr}{1+Rd}\left(-\frac{Nb\Pr}{1+Rd}a^2 - \frac{Nt\Pr}{1+Rd}a^2\right)^2\frac{y^2}{2} + \ldots$$

(32)

$$\gamma = ay + \frac{Nt\Pr}{1+Rd}a^2\frac{y^2}{2} + \frac{Nt^2\Pr}{Nb(1+Rd)}a^2\frac{y^2}{2} - \frac{NbNt\Pr^2}{(1+Rd)^2}a^3\frac{y^3}{6} - \frac{Nt^2\Pr^2}{(1+Rd)^2}a^3\frac{y^3}{6} +$$
$$\frac{Nt^2\Pr}{Nb(1+Rd)}\left(-\frac{Nb\Pr}{1+Rd}a^2 - \frac{Nt\Pr}{1+Rd}a^2\right)\frac{y^2}{2} + \ldots\ldots$$
(33)

$$\varphi = ay + \frac{N_{CT}Nb\Pr}{1+Rd}a^2\frac{y^2}{2} + \frac{N_{CT}Nt\Pr}{1+Rd}a^2\frac{y^2}{2} - \frac{N_{CT}Nb^2\Pr^2}{(1+Rd)^2}a^3\frac{y^3}{6} - \frac{N_{CT}NbNt\Pr^2}{(1+Rd)^2}a^3\frac{y^3}{6}$$
$$+ \frac{N_{CT}Nt\Pr}{(1+Rd)}\left(-\frac{Nb\Pr}{1+Rd}a^2 - \frac{Nt\Pr}{1+Rd}a^2\right)\frac{y^2}{2} + \ldots\ldots$$

(34)

$$\psi = ay^2 - Gr_T a\frac{y^3}{6} - Gr_C a\frac{y^3}{6} + Gr_F a\frac{y^3}{6} + Gr_T \frac{Nb\Pr}{1+Rd}a^2\frac{y^4}{24} + Gr_T \frac{Nt\Pr}{1+Rd}a^2\frac{y^4}{24} + \ldots\ldots$$

(35)

The volume flow rate is given by

$$Q = \int_0^h u\,dy.$$

(36)

Integrating equation (36) and after manipulating we get,

$$\frac{\partial p}{\partial x} = 2Q - a + Gr_T a\frac{1}{9} + Gr_C a\frac{1}{9} - Gr_F a\frac{1}{9} - \frac{Nb\Pr}{1+Rd}Gr_T a^2\frac{1}{192} - \frac{Nt\Pr}{1+Rd}Gr_T a^2\frac{1}{192}.$$

(37)

The volume flow rate in fixed frame is given by

$$Q = \int_0^h (u+1)dy,$$

(38)

$$Q = q + h.$$ (39)

Averaging volume flow rate

$$\bar{Q} = \int_0^1 Q\,dt = \int_0^1 (q+h)dt.$$

(40)

This implies

$$\bar{Q} = q + 1 = 1 + Q - h.$$ (41)

From equation (37) and (41) the pressure gradient is expressed in terms of averaged flow rate as

$$\frac{\partial p}{\partial x} = 2(\bar{Q} - 1 + h) - a + Gr_T a\frac{1}{9} + Gr_C a\frac{1}{9} - Gr_F a\frac{1}{9} - \frac{Nb\Pr}{1+Rd}Gr_T a^2\frac{1}{192} - \frac{Nt\Pr}{1+Rd}Gr_T a^2\frac{1}{192}.$$

(42)

The pressure rise across one wavelength ($\Delta p$) is computed using

$$\Delta p = \int_0^1 \frac{\partial p}{\partial x}dx.$$

(43)

We get,

$$\Delta p = 2(\bar{Q}-1+h) - a + Gr_T a \frac{1}{9} + Gr_C a \frac{1}{9} - Gr_F a \frac{1}{9} - \frac{Nb\,\Pr}{1+Rd} Gr_T a^2 \frac{1}{192} - \frac{Nt\,\Pr}{1+Rd} Gr_T a^2 \frac{1}{192}.$$
(44)

## 4. Discussion

Using HPSTM we have solved the nonlinear partial differential equations. We have used symbolic software Mathematica in the present work. Through the codes of the software Mathematica we have obtained the solutions of HPSTM for velocity, pressure rise, temperature, volume fraction of nanoparticles and solutal (species) concentration profiles and graphical results are plotted in Origin Software. This section constitutes the analysis of different values of physical parameters on velocity, pressure rise, temperature, solutal concentration, and nanoparticle volume fraction.

### 4.1 Velocity distribution

The effects of thermal Grashof number $Gr_T$, solutal Grashof number $Gr_C$, and nanoparticle Grashof number $Gr_F$ on velocity profile are representing through the Figures 2 to 4. Opposite behaviour can be observed in both $Gr_T$ and $Gr_C$. The opposite behaviour can be seen in the case of solutal Grashof number $Gr_C$. As thermal Grashof number $Gr_T$ satisfies the relative influence of thermal buoyancy force and viscous hydrodynamic force. Due to which increasing the thermal Grashof number $Gr_T$ the magnitude of the velocity decreases, For higher values of solutal Grashof number $Gr_C$ the velocity profile increases. Figure 4 shows the effect of nanoparticle Grashof number $Gr_F$ on velocity profile. It is observed that the velocity profile decreases with increasing the $Gr_F$.

### 4.2 Pressure rise

Figures 5-7 are plotted to the different physical parameters they are thermal Grashof number $Gr_T$, Solutal Grashof number $Gr_C$ and nanoparticle Grashof number $Gr_F$ on pressure rise $\Delta p$ against flow rate $Q$. We observed that Figure 5 shows the pressure rise for different values of nanoparticle thermal Grashof number $Gr_T$. We noticed that the pressure rise increases with a higher value of Grashof number $Gr_T$. Figures 6 and 7 shows the pressure rise for different values of nanoparticle thermal solute Grashof number $Gr_C$ and Grashof number $Gr_F$. the pressure rise decreases with an increase in solutal Grashof number $Gr_C$ and nanoparticle Grashof number $Gr_F$. Physically, it is valid because concentration of nanoparticles in the fluid increases, which cause decreasing the pressure.

### 4.3 Temperature distribution

Figures 8-10 depicted for different values of Brownian motion parameter $Nb$, thermophoresis parameter and the thermal radiation parameter $Rd$. From Figures 13 and 14, it is observed that there is enhancement in temperature profile for different values of Brownian motion parameter $Nb$ and thermophoresis parameter $Nt$. The increase of Brownian motion parameter $Nb$ cause the random motion of fluid particles that produce more heat, so there is temperature rises in the system, which can be seen in Figure 8, whereas, in Figure 9, the temperature profile increases as the fluid particles are moved away from the cold surface to hot surface by increasing the thermophoresis parameter $Nt$. In Figure 10, the opposite behaviour can be seen, that is, the decay in the temperature profile with an increase of thermal radiation $Rd$. It is due to fact that the thermal radiation is inversely proportional with thermal conduction parameter $K_T$, therefore maximum heat is radiated away from the system, which leads to reduction in the heat conduction of the fluid.

## 4.4 Solutal (species) concentration distribution

The solutal (species) concentration profile is examined by the profile of $Nb$, $Nt$, $N_{CT}$. Effects $Nb$ and $Nt$ are discussed using Figures 11 and 12. From these two figures, we can observe that the solutal (species) concentration profile has the similar behaviour on both $Nb$ and $Nt$, and solutal (species) concentration decreases with an increasing of $N_{CT}$ we can observe this in figure 13.

## 4.5 Nanoparticle volume fraction distribution

The nanoparticle concentration visualizations for the influences of $Nb$, $Nt$ on peristaltic flow in the presence of nanofluids has been shown in figures 14 and 15. The nanoparticle volume fraction of fluid increases with increase in Brownian motion parameter $Nb$. Because in the case of nanofluids temperature distribution is large, which can lead to the distribution of the system which is shown in figure 14. However, the opposite behaviour is seen in the case of thermophoresis parameter figure 15.

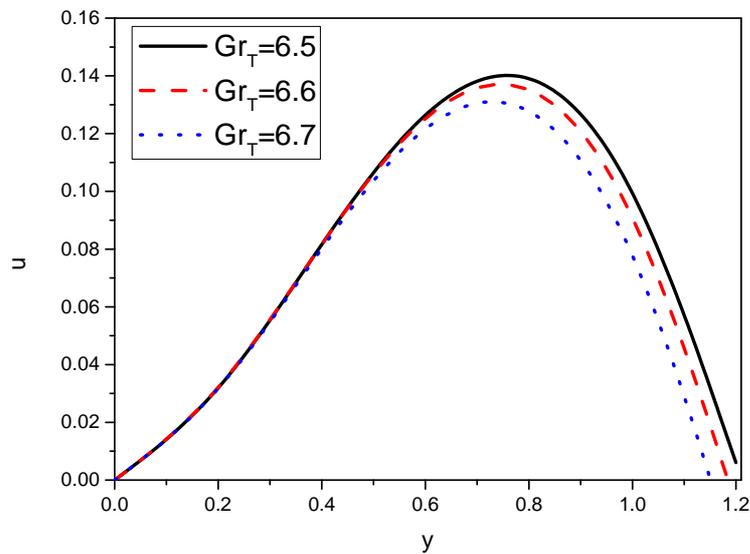

**FIGURE 2** Velocity profile for different values of $Gr_T$
When $Gr_C = 6.5$, $Gr_F = 6.5$, $Nt = 0.5$, $Nb = 0.5$, $\Pr = 1$, $Rd = 0.1$

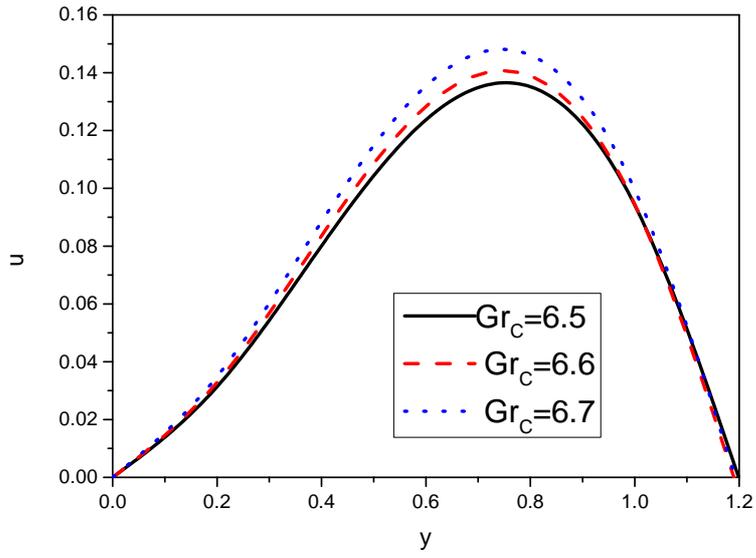

**FIGURE 3** Velocity profile for different values of $Gr_C$
$Gr_T = 6.5$, $Gr_F = 6.5$, $Nt = 0.5$, $Nb = 0.5$, $\Pr = 1$, $Rd = 0.1$

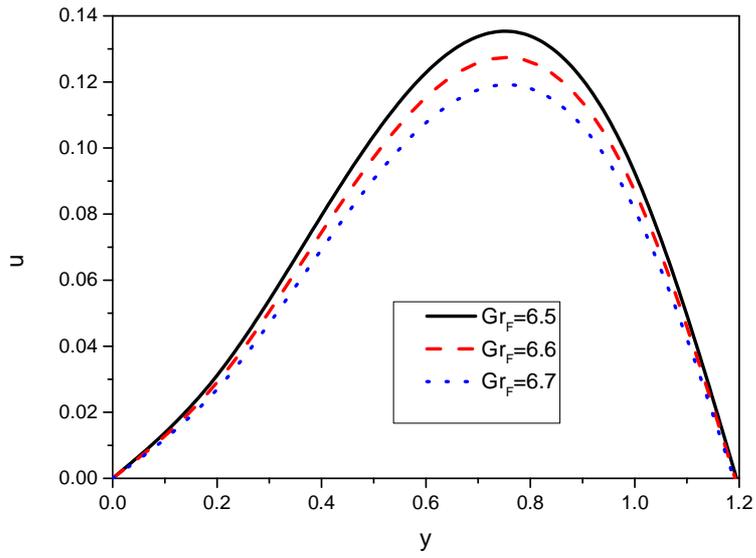

**FIGURE 4** Velocity profile for different values of $Gr_F$
$Gr_T = 6.5$, $Gr_C = 6.5$, $Nt = 0.5$, $Nb = 0.5$, $\Pr = 1$, $Rd = 0.1$

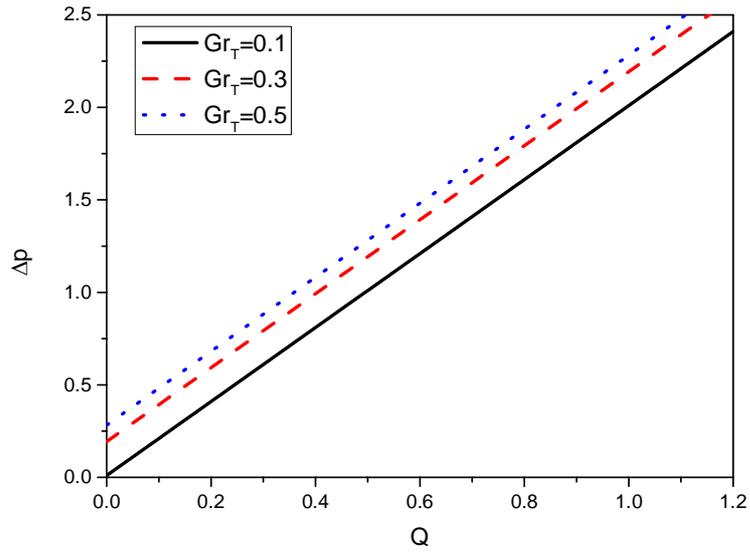

**FIGURE 5** Pressure rise for various values of $Gr_T$ when $Gr_C = 0.8$, $Gr_F = 0.8$, $Nt = 0.6$, $Nb = 0.6$, $\Pr = 7.0$, $Rd = 0.5$

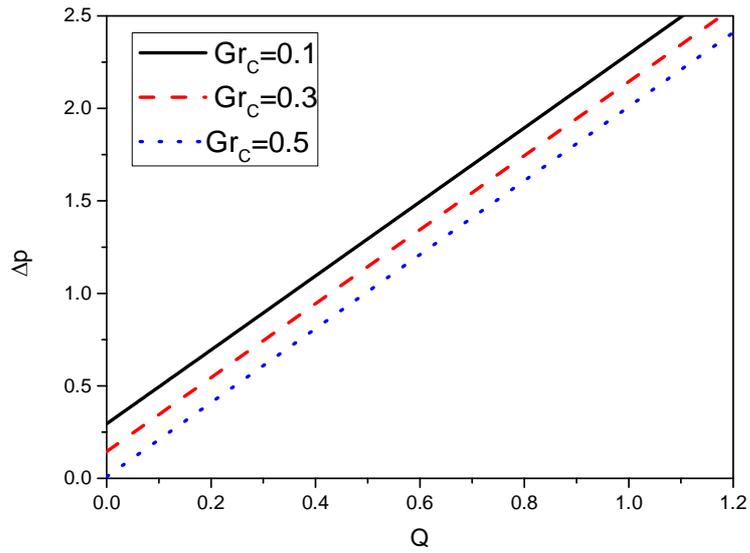

**FIGURE 6** Pressure rise for various values of $Gr_C$ when $Gr_T = 0.8$, $Gr_F = 0.8$, $Nt = 0.6$, $Nb = 0.6$, $\Pr = 7.0$, $Rd = 0.5$

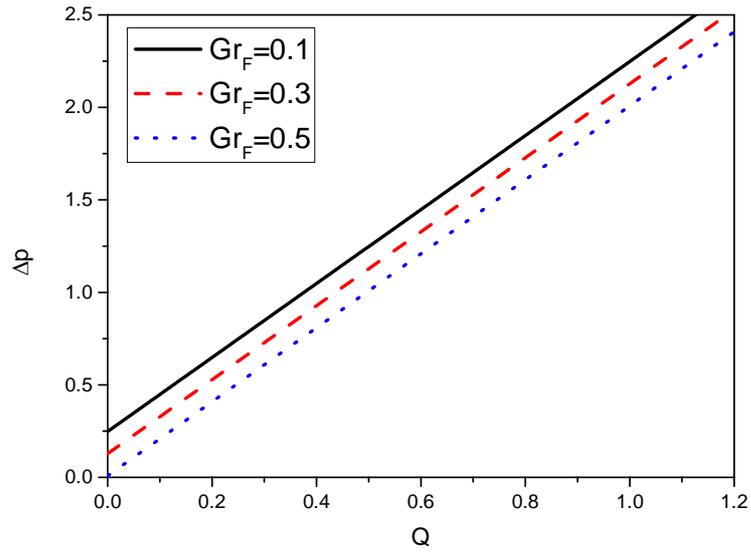

**FIGURE 7** Pressure rise for various values of $Gr_F$
when $Gr_T = 0.8$, $Gr_C = 0.8$, $Nt = 0.6$, $Nb = 0.6$, $\Pr = 7.0$, $Rd = 0.5$

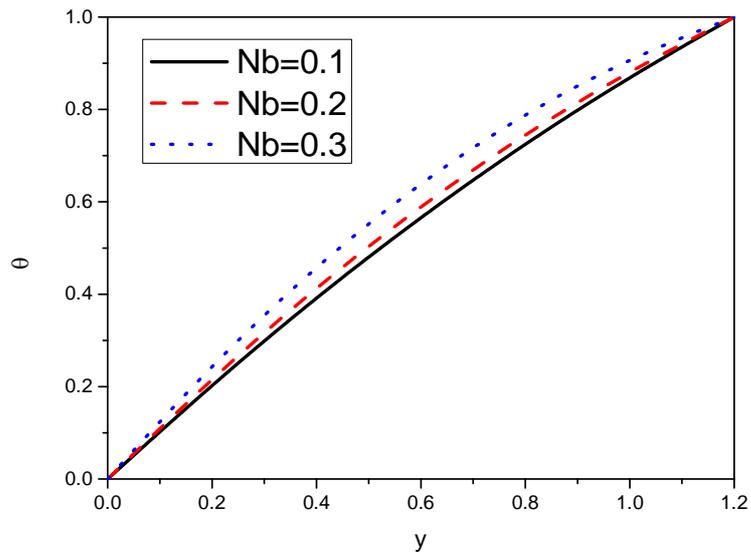

**FIGURE 8** Temperature profile for different values of $Nb$
When $Nt = 0.5$, $\Pr = 1$, and $Rd = 0.9$

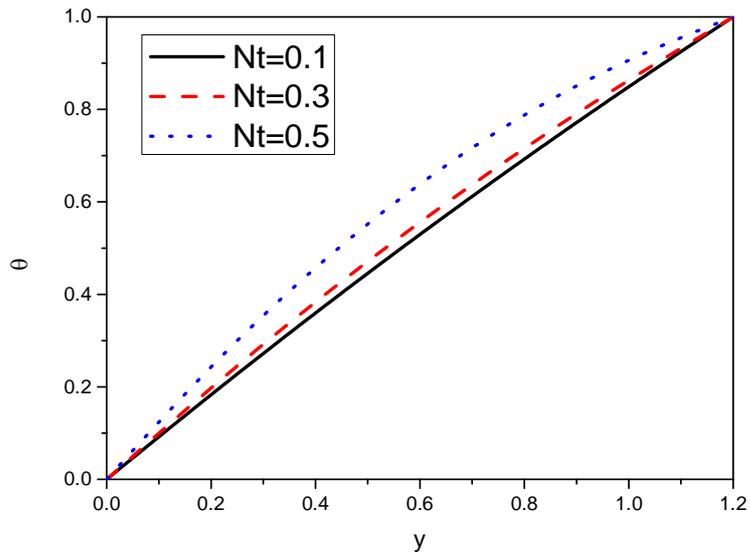

**FIGURE 9** Temperature profile for different values of $Nt$
When $Nb = 0.3$, $\Pr = 1$, and $Rd = 0.9$

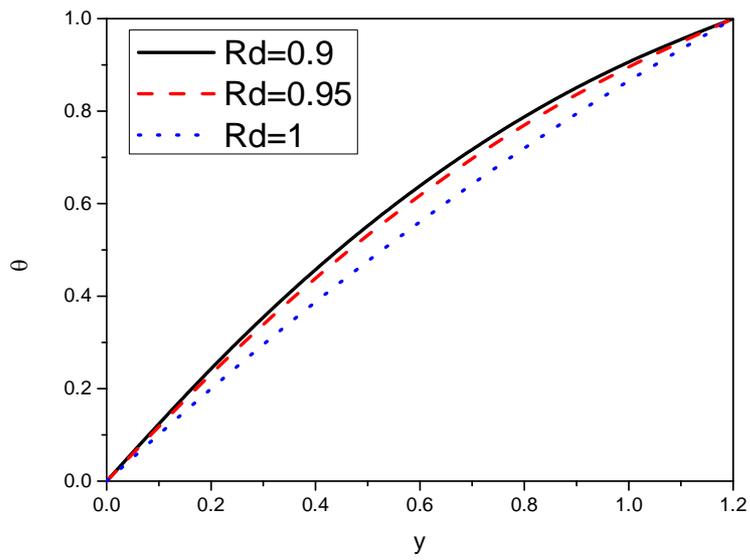

**FIGURE 10** Temperature profile for different values of $Rd$
When $Nb = 0.3$, $\Pr = 1$, and $Nt = 0.5$

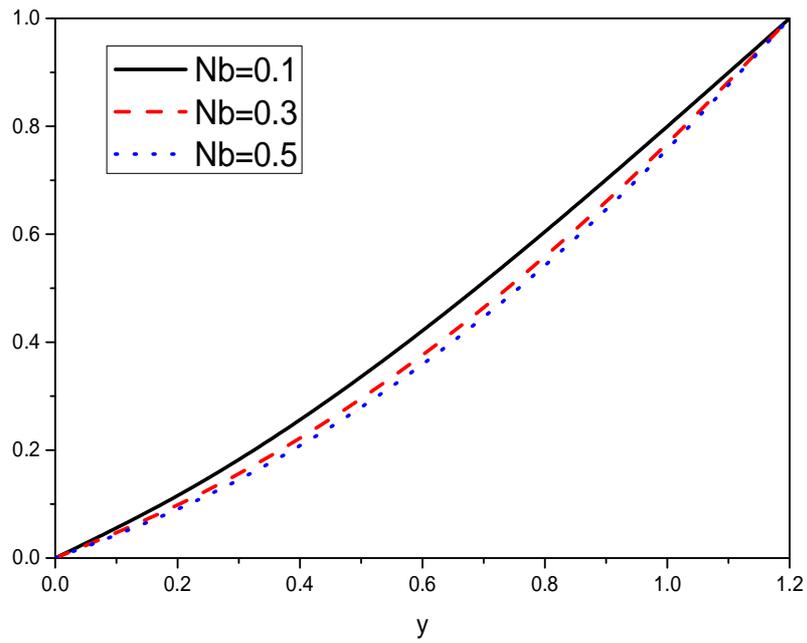

**FIGURE 11** Solutal (species) concentration profile for different values of $Nb$
When $N_{CT}=0.8$, $Nt=0.1$, $\Pr=7.0$, and $Rd=0.1$

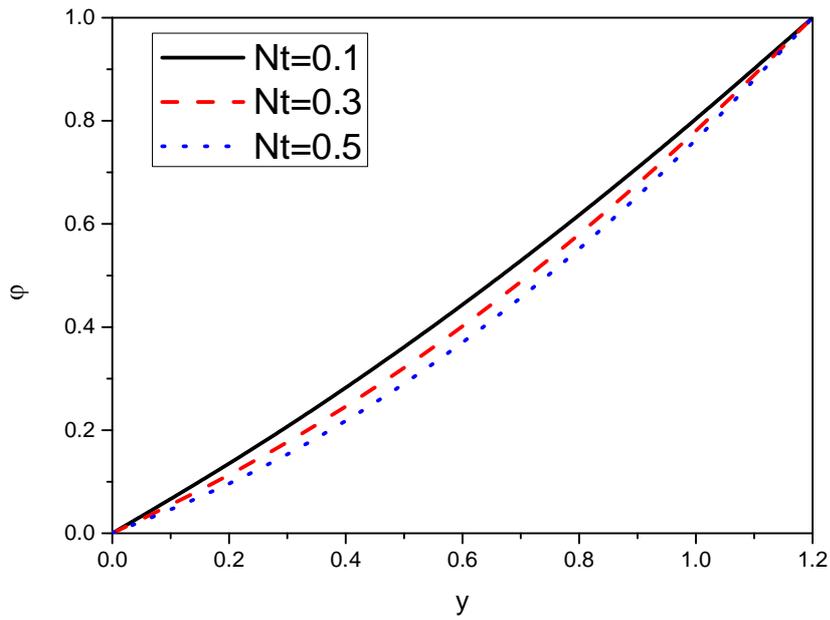

**FIGURE 12** Solutal (species) concentration profile for different values of $Nt$
When $N_{CT}=0.8$, $Nb=0.1$, $\Pr=7.0$, and $Rd=0.1$

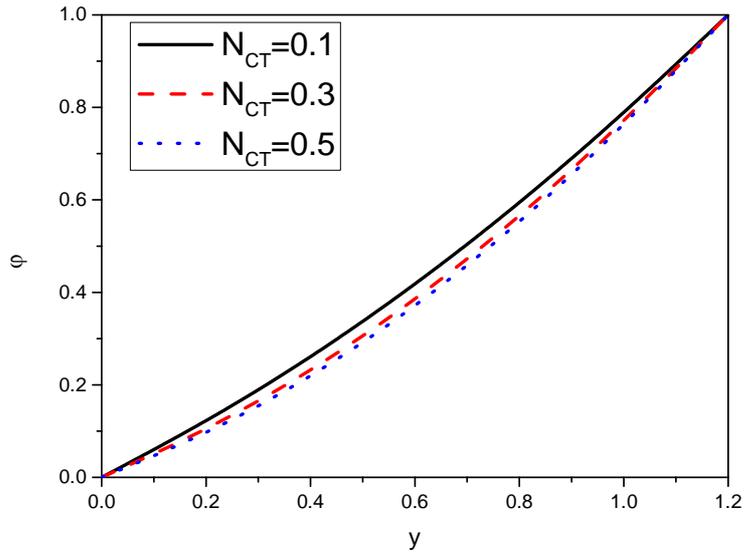

**FIGURE 13** Solutal (species) concentration profile for different values of $N_{CT}$
When $Nt = 0.6$, $Nb = 0.1$, $\Pr = 7.0$, and $Rd = 0.1$

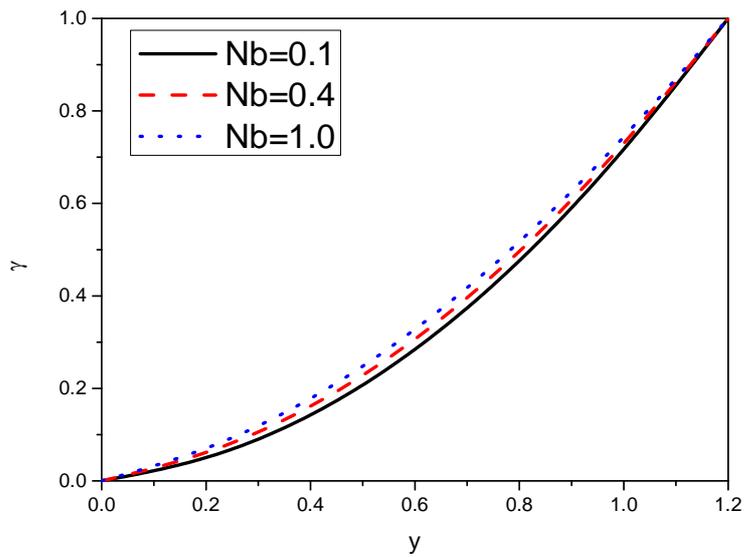

**FIGURE 14** Nanoparticle volume fraction profile for different values of $Nb$
When $Nt = 0.5$, $\Pr = 7.0$, and $Rd = 0.1$

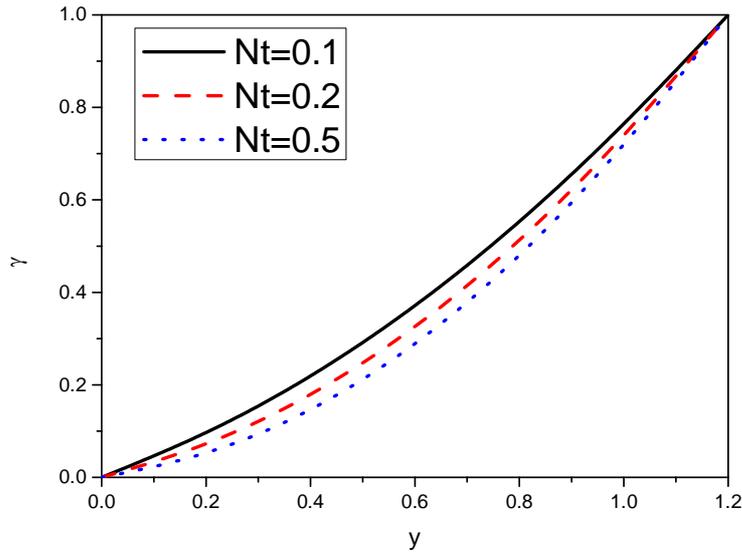

**FIGURE 15** Nanoparticle volume fraction profile for different values of $Nt$
When $Nb = 0.1$, $\Pr = 7.0$, and $Rd = 0.1$

## 5 Conclusion

This research work investigates the double diffusion on peristaltic flow of nanofluid in the presence of porous mediumand thermal radiation through nonuniform channel. The results of the present method are in excellent agreement with the HAM [16]. However, it is worth mentioning that the HPSTM finds the solution without any initial guess or auxiliary linear operator andavoids the rounf-off errors, and it is capable of reducing the volume of the computational work as compared to the HAM while still maintaining the high accuracy of the numerical result and the size reduction amounts to an improvement of the performance of the approach.

The important observations are listed below.
- The present work has potential in biomedical, engineering, and industrial applications.
- The behaviour of relaxation to retardation time on velocity and pressure rise are opposite.
- Opposite behaviours on temperature profile is noted for $Nb$ and $Rd$.
- Behaviour of $Gr_F$ and $Gr_C$ on pressure rise is too similar, that is, increases for $N_{TC}$ and $N_{CT}$ slightly enhanced the temperature of the wall surface.
- Opposite behaviour of $Gr_T$ and $Gr_C$ on velocity profile and pressure rise profile.
- The behaviour of $Gr_T$ and $Gr_C$ on pressure rise are similar.
- The similar behaviour of $Nb$, $Nt$, $N_{CT}$ on solutal (species) concentration profile. Here, solutal concentration profile enhanced with higher values of $Nb$, $Nt$, $N_{CT}$.
- Opposite behaviour of $Nt$ and $Nb$ on nanoparticle volume fraction profile.

**Nomenclature**
$t'$ time (seconds)
$p'$ pressure in fixed frame (Pa)
$h_1'$ right wall

$h_2'$ left wall

$Rd$ Reynolds number

$u'$, $v'$ velocity components (m/s)

$Pr$ Prandtl number

$T'$ temperature (K)

$\varphi'$ nanoparticle volume fraction

$D_B$ Brownian diffusion coefficient (m2/s)

$Gr_F$ nanoparticle Grashof number

$T_m$ fluid mean temperature (K)

$Gr_T$ thermal Grashof number

$F'$ solutal species concentration

$N_{CT}$ Soret parameter

$Nb$ Brownian motion parameter

$Gr_C$ solutalGrashof number